\documentstyle[eqsecnum,aps]{revtex}

\begin{document}
\title{Relativistic theory of the above-threshold multiphoton ionization of\
hydrogen-like\ atoms\ in the ultrastrong laser fields. }
\author{H.K. Avetissian, A.G. Markossian, and G.F. Mkrtchian}
\address{Plasma Physics Laboratory, Department of Theoretical Physics, Yerevan State\\
University, \\
1, A. Manukian St.1, 375049, Yerevan, Armenia\\
Fax: (3742) 151-087 Electronic addresses: avetissian@ysu.am and\\
tesakfiz@sun.ysu.am}
\maketitle

\begin{abstract}
The relativistic theory of above-threshold ionization (ATI) of hydrogen-like
atoms in ultrastrong radiation fields, taking into account the photoelectron
induced rescattering in the continuum spectrum is developed. It is shown
that the contribution of the latter in the multiphoton ionization
probability even in the Born approximation by Coulomb field is of the order
of ATI probability in the scope of Keldysh-Faisal-Reiss ansatz. 
\end{abstract}


\section{Introduction}

The increasing interest to the process of multiphoton above threshold
ionization (ATI) of atoms in superintense laser fields particularly is
conditioned by the problem of high harmonic generation and short wave
coherent radiation implementation via multiphoton bound-free transitions
through free continuum spectrum (one of the possible version of a free
electron X-ray laser). During the last two decades numerous investigations
have been carried out to study ATI of atoms both theoretically and
experimentally and many review papers (see, e.g., \cite{1}-\cite{10} ) and
monographs \cite{11}-\cite{15} are devoted to this problem.

The main ansatz in the nonrelativistic theory of multiphoton ionization of
atoms in strong electromagnetic (EM) radiation fields is the Keldysh one 
\cite{16} (further called Keldysh-Faisal-Reiss ansatz \cite{17}, \cite{18}).
The advantage of this approach is that it leads in a very simple way to
reveal some of the main qualitative features of the photoelectron energy
spectrum in ATI experiments (\cite{a}, \cite{b} and \cite{c}). Within the
scope of this ansatz the photoelectron rescattering in the field of atomic
remainder is neglected. In order to cover this gap attempts have been made
to describe the photoelectron final state by ''Coulomb-Volkov'' wave
function that is a product of the Coulomb wave function of elastic
scattering and a wave function of electron in the EM wave field \cite{19}-%
\cite{25}. This wave function results in the factorization of the
probability of multiphoton ionization and restricts both the frequency (low
frequency approximation) and intensity of the wave. The use of another
ansatz for definition of multiphoton ionization probabilities \cite{26}
should also be noted.

The description of the photoelectron final state taking into account the
stimulated bremsstrahlung (SB) at the photoelectron scattering on the
electrostatic potential of the ionized atom in the presence of strong EM
radiation field (induced free-free transitions) still remains as one of the
main problems for the ATI process. Moreover, the definition of dynamic wave
function of an electron in SB process already is problematic, so that the
main results concerning to multiphoton SB probabilities have been found
trough S-matrix formalism for ''free-free'' (over electrostatic potential)
transitions in the Born approximation between Volkov states in EM wave \cite
{27}. Although, in a lot of cases when the condition of the Born
approximation is broken, the scattering process is described in
low-frequency \cite{28}, \cite{29} or eikonal \cite{30} approximations.
Though the Born and low-frequency approximations in SB process are
applicable for describing free-free transitions in high intensity radiation
fields, but they don't take into account the mutual influence of the
scattering and the radiation fields [i.e. the probability of SB is
factorized by elastic scattering and photon emission or absorption
processes]. What concerns to the eikonal approximation in SB process it is
not applicable beyond the interaction region [ $z\ll pa^2/\hbar $, where $z$
is the coordinate along the direction of initial momentum $\overrightarrow{p}
$ of the particle, $a$ is the range of the interaction region, and $\hbar $
is the Plank constant]. The description of the electron eigenstates in SB
process behind of the scope of these approximations has been made in \cite
{31}, developing a generalized eikonal approximation (GEA). The obtained GEA
wave function enables to leave the framework of ordinary eikonal
approximation and abandons from the restriction $z\ll pa^2/\hbar $. Besides,
such wave function simultaneously takes into account the influence of both
the scattering and radiation fields on the particle state. So, to determine
the multiphoton probabilities of above threshold ionization of an atom it
should be known the wave function of the ejected photoelectron in SB process
with more accuracy. On the other hand, in the current superintense laser
fields the state of electron becomes relativistic already at the distances $%
l\ll \lambda $ ($\lambda $ is the wavelength of a laser radiation)
independent on its initial state. Hence, the problem of ATI of atoms with
the photodetached electron SB process demands a relativistic consideration.

The relativistic generalization of multiphoton SB in the first Born and
eikonal approximations have been made in the papers \cite{32}, \cite{33} and 
\cite{34} respectively. In \cite{35} on the base of the solution of the
Dirac equation the GEA approximation \cite{31} has been developed for
relativistic scattering theory in the arbitrary electrostatic and plane EM
wave fields, including both the Born and eikonal approximations in
corresponding limits and describing the spin interaction as well. Such a
wave function allows us to describe the final state of the photoelectron
with more accuracy in the ATI process of atoms.

The relativistic description of multiphoton ATI of hydrogen-like atoms for
high-intensity laser fields taking into account the spin interaction has
been developed analytically in the papers \cite{36}-\cite{38} with an
approximation where the stimulated bremsstrahlung of the emergent electron
is neglected. The relativistic consideration of ATI is important as it is
generally assumed that the problem of stabilization of atoms in ultraintense
laser fields must be solved within the framework of relativistic theory \cite
{39}.{\bf \ } From this point of view some attempts have been made to solve
analytically the Klein-Gordon equation \cite{40}, \cite{41} or numerically
the Dirac equation \cite{39}, \cite{42} in fields of a static potential and
monochromatic EM wave [using various model potentials of one or two
dimensions and various approximations]. In the papers \cite{43}-\cite{45}
the relativistic corrections to the nonrelativistic results have been given.

Note, that at the present time an analytic formulas for these probabilities,
taking into account the photoelectron rescattering, are unknown even in the
first Born approximation for the Coulomb scattering field. So, in the
present paper the relativistic probabilities of multiphoton ATI in the limit
of the Born approximation for the photoelectron rescattering, are
calculated. Moreover, it is shown that the neglect of the photoelectron
rescattering in the relativistic domain specially \cite{36}-\cite{38} is
invalid, since the contribution of the electron rescattering process in the
matrix elements of transitions has the same order by a scattering potential
in the Born approximation as the matrix elements of bound-free transitions
for the ATI process.

The organization of the paper is as follows. In Sec. II we present the
multiphoton cross sections of the above threshold ionization of
hydrogen-like atom in ultraintense laser field (with the help of GEA wave
function), taking into account the induced free-free transitions of the
ejected photoelectron in the continuum spectrum. Because of much complicated
expressions in GEA the spin interaction is neglected and the ultimate
analytic results for the multiphoton probabilities are performed in the
limit of the first Born approximation by ion (atomic remainder) potential
which we present in Sec. III. In Sec. IV we treat the dependence of ATI
probability on polarization of an EM wave and consider the differencies
between circular and linear polarizations of electromagnetic wave.

\section{The\ ionization probability by relativistic GEA solution of a wave
equation of an electron}

The problem has been reduced to the investigation of the relativistic
exploration of the transition S-matrix formalism utilizing the relativistic
GEA wave function \cite{35} as a wave function of the final state of a
photodetached electron (it has been neglected with the spin interaction in
relativistic GEA wave function). Following the relativistic S-matrix
formalism the bound-free transition amplitude can be written in this
integral form (in natural units $\hbar =c=1$) 
\begin{equation}
T_{i\rightarrow f}=-i\int_{-\infty }^\infty \Psi ^{(-)\dagger \ }(x)\widehat{%
V}\Phi (x)d^4x\text{ ,}  \label{1}
\end{equation}
where $x=(t,\overrightarrow{r})$ is the four-component radius-vector $x^\mu $%
, $\Phi (x)$ is the initial unperturbed bound state of the atomic system and 
$\Psi ^{(-)}(x)$ is the final out-state of an electron in the potential of
atomic remainder and in the field of a plane EM wave ($K^{\dagger }$ is
denotes the complex conjugation of $K$). We assume the EM wave to be
quasimonochromatic and of an arbitrary polarization with the vector
potential 
\begin{equation}
\overrightarrow{A}(\varphi )=A_0(\varphi )\left( \overrightarrow{e}_1\cos
\varphi +\overrightarrow{e}_2\zeta \sin \varphi \right) ;\qquad \varphi
=k\cdot x=\omega t-\overrightarrow{k}{\bf \cdot }\overrightarrow{r},
\label{2}
\end{equation}
where $k=(\omega ,\overrightarrow{k})$ is the four-wave vector, $A_0(\varphi
)$ is the slow varying amplitude of the vector potential of a plane wave, $%
\overrightarrow{e}_1$ and $\overrightarrow{e}_2$ are unit vectors: $%
\overrightarrow{e}_1$ $\perp $ $\overrightarrow{e}_2$ $\perp \overrightarrow{%
k}$ , and $\arctan \zeta $ is the polarization angle.

According to the Klein-Gordon equation the interaction operator is

\begin{equation}  \label{3}
\widehat{V}=-2e\overrightarrow{A}(\varphi )(-i\overrightarrow{\nabla })+e^2%
\overrightarrow{A}^2(\varphi ),
\end{equation}
where $e$ is the electron charge.

The wave function of the final state of the photodetached electron in the
relativistic GEA approximation has the following form (\cite{35}) 
\begin{equation}  \label{4}
\Psi ^{(-)\dagger \ }(x)=\frac 1{\sqrt{2\Pi _0}}F^{\dagger }(x)\exp
[-iS_V(x)],
\end{equation}

The $S_V(x)$ is the action of photoelectron in the field (\ref{2})

\begin{equation}  \label{5}
S_V(x)=\overrightarrow{\Pi }{\bf \cdot }\overrightarrow{r}-\Pi _0t+\alpha
\left( \frac{\overrightarrow{p}}{k\cdot p}\right) \sin [\varphi -\theta (%
\overrightarrow{p})]-\frac Z2(1-\zeta ^2)\sin 2\varphi .
\end{equation}
Here $\Pi =(\Pi _0,\overrightarrow{\Pi })$ is the average four- kinetic
momentum or ''quasimomentum'' of the electron in the plane EM wave field,
which is defining via free electron four-momentum $p=(\varepsilon _0,%
\overrightarrow{p})$ and relative parameter of the wave intensity $Z$ by the
following equation

\begin{equation}
\Pi =p+kZ(1+\zeta ^2);\qquad Z=\frac{e^2\overline{A}_0^2}{4k\cdot p},
\label{6}
\end{equation}
where $\overline{A}_0$ is the averaged value of the amplitude $A_0(\varphi )$%
. The wave function (\ref{4}) is normalized for the one particle in the unit
volume $V=1$.

Including in (\ref{5}) quantity $\alpha \left( \frac{\overrightarrow{p}}{%
k\cdot p}\right) $ is the intensity-dependent amplitude of the electron-wave
interaction and as a function on any three-vector $\overrightarrow{b}$ has
the following definition 
\begin{equation}  \label{7}
\alpha \left( \overrightarrow{b}\right) =e\overline{A}_0\sqrt{\left( 
\overrightarrow{b}\cdot \overrightarrow{e}_1\right) ^2+\zeta ^2\left( 
\overrightarrow{b}\cdot \overrightarrow{e}_2\right) ^2},
\end{equation}
with the phase angle 
\begin{equation}  \label{8}
\theta (\overrightarrow{p})=\arctan \left( \frac{\overrightarrow{p}\cdot 
\overrightarrow{e}_2}{\overrightarrow{p}{\bf \cdot }\overrightarrow{e}_1}%
\zeta \right) .
\end{equation}

The function $F^{\dagger }(x)$ in Eq. (\ref{4}), which describes the impact
of both the scattering and EM radiation fields on the photoelectron state
simultaneously, has the following form \cite{35} 
\[
F^{\dagger }(x)=\exp \left[ \frac 1{4\pi ^3}\sum\limits_{n=-\infty }^\infty
e^{in\varphi }\int \frac{\left\{ \omega \left[ \alpha \left( \frac{%
\overrightarrow{p}}{k\cdot p}\right) D_{1,n}^{\dagger }\left( \theta _1(%
\overrightarrow{q})-\theta (\overrightarrow{p})\right) -Z(1-\zeta
^2)D_{2,n}^{\dagger }\right] -\Pi _0D_n^{\dagger }\right\} }{\overrightarrow{%
q}^2+2\overrightarrow{\Pi }{\bf \cdot }\overrightarrow{q}-2n(k\cdot p-%
\overrightarrow{k}{\bf \cdot }\overrightarrow{q})+i0}\right. 
\]
\begin{equation}
\left. \times \widetilde{U}(\overrightarrow{q})\exp \left[ -i\left\{ 
\overrightarrow{q}\cdot \overrightarrow{r}+\alpha _1{\bf (}\overrightarrow{q}%
{\bf )}\sin \left[ \varphi -\theta _1(\overrightarrow{q})\right] -\alpha _2(%
\overrightarrow{q})\sin 2\varphi +\theta _1(\overrightarrow{q})n\right\} d%
\overrightarrow{q}\right] \right] ,  \label{9}
\end{equation}
where 
\begin{equation}
\widetilde{U}(\overrightarrow{q})=\int U{\bf (}\overrightarrow{r})\exp (-i%
\overrightarrow{q}{\bf \cdot }\overrightarrow{r})d\overrightarrow{r}
\label{10}
\end{equation}
is the Fourier transform of the potential of the atomic remainder and $%
\alpha _1(\overrightarrow{q})$ , $\alpha _2(\overrightarrow{q})$ are dynamic
parameters of the interaction defining by expression 
\begin{equation}
\alpha _1(\overrightarrow{q})=\alpha \left( {\bf (}\overrightarrow{k}\cdot 
\overrightarrow{q}{\bf )}\overrightarrow{p}/k\cdot p+\overrightarrow{q}%
\right) ,\ \alpha _2(\overrightarrow{q})=\frac{\overrightarrow{k}\cdot 
\overrightarrow{q}}{2(k\cdot p-\overrightarrow{k}\cdot \overrightarrow{q}%
{\bf )}}Z(1-\zeta ^2),  \label{11}
\end{equation}
and $\theta _1(\overrightarrow{q})$ is the phase angle 
\begin{equation}
\theta _1(\overrightarrow{q})=\theta \left( {\bf (}\overrightarrow{k}\cdot 
\overrightarrow{q}{\bf )}\overrightarrow{p}/k\cdot p+\overrightarrow{q}%
\right) .  \label{11a}
\end{equation}
The functions $J_n(u,v,\triangle ),$ $D_n,$ $\,D_{1,n}\left( \theta _1(%
\overrightarrow{q})-\theta (\overrightarrow{p})\right) ,$ and $D_{2,n}$ are
defined by the expressions (also see Ref. \cite{35}) 
\begin{equation}
D_n=J_n(\alpha _1(\overrightarrow{q}),-\alpha _2(\overrightarrow{q}),\theta
_1(\overrightarrow{q})),  \label{12}
\end{equation}
\[
D_{1,n}\left( \theta _1(\overrightarrow{q})-\theta (\overrightarrow{p}%
)\right) =\frac 12\left[ J_{n-1}(\alpha _1(\overrightarrow{q}),-\alpha _2(%
\overrightarrow{q}),\theta _1(\overrightarrow{q}))e^{-i\left( \theta _1(%
\overrightarrow{q})-\theta (\overrightarrow{p})\right) }\right. 
\]
\begin{equation}
\left. +J_{n+1}(\alpha _1(\overrightarrow{q}),-\alpha _2(\overrightarrow{q}%
),\theta _1(\overrightarrow{q}))e^{i\left( \theta _1(\overrightarrow{q}%
)-\theta (\overrightarrow{p})\right) }\right] ,  \label{13}
\end{equation}
and 
\[
D_{2,n}=\frac 12\left[ J_{n-2}(\alpha _1(\overrightarrow{q}),-\alpha _2(%
\overrightarrow{q}),\theta _1(\overrightarrow{q}))e^{-i2\theta _1(%
\overrightarrow{q})}\right. 
\]
\begin{equation}
+J_{n+2}(\alpha _1(\overrightarrow{q}),-\alpha _2(\overrightarrow{q}),\theta
_1(\overrightarrow{q}))e^{i2\theta _1(\overrightarrow{q})}.  \label{14}
\end{equation}

In the denominator of the integral in expression (\ref{9}) $+i0$ is an
imaginary infinitesimal, which shows how the path around the pole in the
integrand should be chosen to obtain a certain asymptotic behavior of the
wave function, i.e., the ingoing spherical wave{\bf \ [}to determine that%
{\bf \ }one must be passed to the limit of the Born approximation at $%
\overrightarrow{A}{\bf (}\varphi {\bf )}=\overrightarrow{0}$].

Since we consider the ATI problem for hydrogen-like atoms ( $Z_a\ll 137$),
the initial velocities of atomic electrons are nonrelativistic and as a
initial-state wave function $\Phi $ in the transition amplitude (\ref{1})
will be taken a stationary wave function of hydrogen-like atom bound state
in nonrelativistic limit 
\begin{equation}
\Phi \left( \overrightarrow{r},t\right) =\frac 1{\sqrt{2m}}\Phi _0\left( 
\overrightarrow{r}\right) \exp \left( -i\varepsilon _0t\right) ,\text{
\qquad }\varepsilon _{0\text{ }}=m-E_B,  \label{20}
\end{equation}
where $E_B\succ 0$ is the binding energy of the valence electron in the atom 
\begin{equation}
2mE_B=a^{-2}.  \label{20b}
\end{equation}

Concerning the relativism of the photoelectron final state in strong EM
field it is followed to mention that at the wave intensities already $\xi
\sim 10^{-1}$, where

\begin{equation}  \label{20a}
\xi =\frac{e\overline{A}_0}m
\end{equation}
is the relativistic invariant parameter of the wave intensity, relativistic
effects become observable and the final state of the photoelectron should be
described in the scope of relativistic theory. Moreover, at the currently
available laser intensities $\xi \succ 1$\ (even $\xi \gg 1$){\bf \ }a free
electron becomes essentially relativistic already at the distances smaller
than one wavelength. On the other hand, in such fields becomes actual the
production of electron-positron pairs from intense photon field on the
electrostatic potential of atomic remainder through multiphoton channels.
However, we can calculate separately the ATI probability in superstrong
laser fields without restricting intensities by the threshold value of
multiphoton pairs production ($\xi \simeq 2$; see \cite{46} and \cite{47})
since those are independent processes.

Since $\widehat{V}$ is a Hermitian operator, the transition amplitude (\ref
{1}) can be written in the form 
\begin{equation}  \label{21}
T_{i\rightarrow f}=-i\int_{-\infty }^\infty \Phi (x)\widehat{V}^{\dagger
}(x)\Psi ^{(-)\ \dagger }(x)d^4x.
\end{equation}
To integrate this expression it is convenient to turn from variables $t,%
\overrightarrow{r}$ to $\varphi $, $\overrightarrow{\eta }$[see (\ref{2})] 
\begin{equation}  \label{22}
T_{i\rightarrow f}=-\frac i\omega \int_{-\infty }^\infty \Phi (\varphi ,%
\overrightarrow{\eta })\widehat{V}^{\dagger }(\varphi ,\overrightarrow{\eta }%
)\Psi ^{(-)\ \dagger }(\varphi ,\overrightarrow{\eta })d\varphi d%
\overrightarrow{\eta }.
\end{equation}
and make a Fourier transformation of the function $F^{\dagger }(x)$ over
variable $\varphi $ 
\begin{equation}  \label{23}
F^{\dagger }(\varphi ,\overrightarrow{\eta })=\sum_{l=-\infty }^\infty 
\widetilde{F_l}(\overrightarrow{\eta })\exp (-il\varphi ){\bf ,}
\end{equation}
\begin{equation}  \label{24}
\widetilde{F_l}(\overrightarrow{\eta })=\frac 1{2\pi }\int_{-\pi }^\pi
F(\varphi ,\overrightarrow{\eta })\exp (il\varphi )d\varphi {\bf .}
\end{equation}
Then with the help of Eqs. (\ref{2})-(\ref{14}), (\ref{20}) [using as well
Eq. (\ref{A3})] and taking into account the Lorentz condition for the plane
wave field $\overrightarrow{k}\cdot \overrightarrow{A}(\varphi )=0,$ we can
accomplish the integration over the variable $\varphi $ in Eq. (\ref{22}).
After a simple transformation with the help of the formula (\ref{A4}) we
obtain the next expression for the transition amplitude 
\[
T_{i\rightarrow f}=\frac{i2\pi (k\cdot p)}{\omega \sqrt{m\Pi _0}}%
\sum_{L,l=-\infty }^\infty \left\{ \left( L-Z(1+\zeta ^2)\right) \widetilde{%
\Phi }_l\left( \overrightarrow{g}\right) J_L\left( \alpha \left[ \frac{%
\overrightarrow{p}}{k\cdot p}\right] ,-\frac Z2(1-\zeta ^2),\theta (%
\overrightarrow{p})\right) e^{iL\theta (\overrightarrow{p})}\right. 
\]
\[
\times \delta \left( \frac{\Pi _0-\varepsilon _0}\omega -L-l\right) 
\]
\[
+2\sum_{n=-\infty }^\infty \int \frac{d\overrightarrow{q}}{(2\pi )^3}%
\widetilde{\Phi }_l\left( \overrightarrow{g}+\overrightarrow{q}\right) 
\widetilde{U}(\overrightarrow{q}) 
\]
\[
\times \alpha \left( \frac{\overrightarrow{q}}{k\cdot p}\right)
C_{1,L}^{\dagger }\left( \theta (\overrightarrow{p}+\overrightarrow{q}%
)-\theta (\overrightarrow{q})\right) e^{-in\theta _1(\overrightarrow{q}%
)+iL\theta (\overrightarrow{p}+\overrightarrow{q})} 
\]
\[
\times \frac{\left\{ \omega \left[ \alpha \left( \frac{\overrightarrow{p}}{%
k\cdot p}\right) D_{1,n}^{\dagger }\left( \theta _1(\overrightarrow{q}%
)-\theta (\overrightarrow{p})\right) -Z(1-\zeta ^2)D_{2,n}^{\dagger }\right]
-\Pi _0D_n^{\dagger }\right\} }{\overrightarrow{q}^2+2\overrightarrow{\Pi }%
{\bf \cdot }\overrightarrow{q}-2n(k\cdot p-\overrightarrow{k}{\bf \cdot }%
\overrightarrow{q})+i0} 
\]
\begin{equation}  \label{26}
\times \left. \delta \left( \frac{\Pi _0-\varepsilon _0}\omega -L-l+n\right)
\right\} ,
\end{equation}
where $\overrightarrow{g}$ is the three-vector 
\begin{equation}  \label{27}
\overrightarrow{g}=\overrightarrow{p}-\frac{(\varepsilon -\varepsilon _0)%
\overrightarrow{k}}\omega .
\end{equation}
and the function $\widetilde{\Phi }_l\left( \overrightarrow{b}\right) $ is
the Fourier transform of $\Phi _l(\overrightarrow{\eta })\equiv \Phi (%
\overrightarrow{\eta })\widetilde{F_l}(\overrightarrow{\eta })$ and as a
function on any three-vector $\overrightarrow{b}$is defined by the formula (%
\ref{10}), and 
\[
C_{1,n}\left( \theta (\overrightarrow{p}+\overrightarrow{q})-\theta (%
\overrightarrow{q})\right) 
\]
\[
=\frac 12\left[ J_{n-1}(\alpha _1(\overrightarrow{p}+\overrightarrow{q}),-%
\frac{Z_1}2(1-\zeta ^2),\theta (\overrightarrow{p}+\overrightarrow{q}%
))e^{-i\left( \theta (\overrightarrow{p}+\overrightarrow{q})-\theta (%
\overrightarrow{q})\right) }\right. 
\]
\begin{equation}  \label{28a}
\left. +J_{n+1}(\alpha _1(\overrightarrow{q}),-\frac{Z_1}2(1-\zeta
^2),\theta (\overrightarrow{p}+\overrightarrow{q}))e^{i\left( \theta (%
\overrightarrow{p}+\overrightarrow{q})-\theta (\overrightarrow{q})\right)
}\right] ,
\end{equation}
where the parameters $\alpha \left( \frac{\overrightarrow{p}+\overrightarrow{%
q}}{k\cdot p-\overrightarrow{k}\cdot \overrightarrow{q}}\right) $, $\theta (%
\overrightarrow{p}+\overrightarrow{q})$ are determined by the expressions (%
\ref{7}) and (\ref{8}), and 
\begin{equation}  \label{28b}
Z_1=\frac{e^2\overline{A}_0^2}{4(k\cdot p-\overrightarrow{k}\cdot 
\overrightarrow{q})}.
\end{equation}

Using the general conservation law of considering process the probability
amplitude of the above-threshold ionization in concluding form can be
presented in this ultimate form 
\[
T_{i\rightarrow f}=\frac{i2\pi (k\cdot p)}{\sqrt{m\Pi _0}}\sum_{N,l=-\infty
}^\infty \left\{ (N+l-Z(1+\zeta ^2))\widetilde{\Phi }_l\left( 
\overrightarrow{g}\right) \right. 
\]
\[
\times J_{N-l}\left( \alpha \left( \frac{\overrightarrow{p}}{k\cdot p}%
\right) ,-\frac Z2(1-\zeta ^2),\theta (\overrightarrow{p})\right)
e^{i(N-l)\theta (\overrightarrow{p})} 
\]
\[
+2\sum_{n=-\infty }^\infty \int \frac{d\overrightarrow{q}}{(2\pi )^3}%
\widetilde{\Phi }_l\left( \overrightarrow{g}+\overrightarrow{q}\right) 
\widetilde{U}(\overrightarrow{q}) 
\]
\[
\times \alpha \left( \frac{\overrightarrow{q}}{k\cdot p}\right)
C_{1,N-l+n}^{\dagger }\left( \theta (\overrightarrow{p}+\overrightarrow{q}%
)-\theta (\overrightarrow{q})\right) e^{-in\theta _1(\overrightarrow{q}%
)+i(N-l+n)\theta (\overrightarrow{p}+\overrightarrow{q})} 
\]
\[
\left. \times \frac{\left\{ \omega \left[ \alpha \left( \frac{%
\overrightarrow{p}}{k\cdot p}\right) D_{1,n}^{\dagger }\left( \theta _1(%
\overrightarrow{q})-\theta (\overrightarrow{p})\right) -Z(1-\zeta
^2)D_{2,n}^{\dagger }\right] -\Pi _0D_n^{\dagger }\right\} }{\overrightarrow{%
q}^2+2\overrightarrow{\Pi }{\bf \cdot }\overrightarrow{q}-2n(k\cdot p-%
\overrightarrow{k}{\bf \cdot }\overrightarrow{q})+i0}\right\} 
\]
\begin{equation}  \label{29}
\times \delta \left( \Pi _0-\varepsilon _0-\omega N\right) .
\end{equation}

The differential probability of ATI process per unit time in the phase space 
$d\overrightarrow{\Pi }/\left( 2\pi \right) ^3$ (space volume $V=1$ in
accordance with normalization of electron wave function) taking into account
the all final states of photoelectron with quasimomenta in the interval $%
\overrightarrow{\Pi },$ $\overrightarrow{\Pi }+d\overrightarrow{\Pi }$ is 
\[
dW_{i\rightarrow f}=w_{i\rightarrow f}\frac{d\overrightarrow{\Pi }}{\left(
2\pi \right) ^3} 
\]
\begin{equation}  \label{30}
=w_{i\rightarrow f}\sqrt{\Pi _0^2-m_{*}^2}\Pi _0d\Pi _0\frac{d\Omega }{(2\pi
)^3},
\end{equation}
where $d\Omega $ is the differential solid angle and 
\begin{equation}  \label{31}
m_{*}=\sqrt{\Pi _0^2-\overrightarrow{\Pi }^2}=\sqrt{m^2+e^2\overline{A}_0^2%
\frac{(1+\zeta ^2)}2}
\end{equation}
is the ''effective mass'' of the relativistic electron in the EM wave field.

Including in Eq.(\ref{30}) the transition probability per unit time $%
w_{i\rightarrow f}$ has the general definition 
\begin{equation}  \label{32}
w_{i\rightarrow f}=\lim _{t\rightarrow \infty }\frac 1t\left|
T_{i\rightarrow f}\right| ^2
\end{equation}

\section{The relativistic Born approximation by the potential of atomic
remainder for hydrogen-like atom ionization}

The impact of rescattering effect on the ATI process is more transparent in
the limit of the Born approximation by the scattering potential. The latter
takes place if the corresponding part of the action $S_1\left( 
\overrightarrow{r}{\bf ,}t\right) $ in the GEA wave function (\ref{4}),
i.e., the exponent of the function $F^{\dagger }(x)$ ($F^{\dagger }\left(
x\right) =\exp \left[ iS_1(\overrightarrow{r}{\bf ,}t)\right] $) is enough
small (see Ref. \cite{35}):

\begin{equation}  \label{33}
\left| S_1\left( \overrightarrow{r}{\bf ,}t\right) \right| \ll 1\text{ }.%
\text{ }
\end{equation}

Expanding Eq. (\ref{29}) into the series and keeping only the terms to the
first order over $U(\overrightarrow{r}),$ after a simple transformation,
utilizing Eqs. (\ref{A2}), (\ref{A3}) and (\ref{A4}), we obtain 
\[
T_{i\rightarrow f}=\frac{i2\pi }{\sqrt{m\Pi _0}}\sum_{N=-\infty }^\infty
\left\{ \left[ N-Z(1+\zeta ^2)\right] (k\cdot p)\widetilde{\Phi }\left( 
\overrightarrow{g}\right) e^{iN\theta (\overrightarrow{p})}J_N\left( \alpha
\left( \frac{\overrightarrow{p}}{k\cdot p}\right) ,-\frac Z2(1-\zeta
^2),\theta (\overrightarrow{p})\right) \right. 
\]
\[
+2\sum_{n=-\infty }^\infty \int \frac{d\overrightarrow{q}}{(2\pi )^3}\left[
N+n-Z_1(1+\zeta ^2)\right] \left( k\cdot p-\overrightarrow{k}\cdot 
\overrightarrow{q}\right) \widetilde{\Phi }\left( \overrightarrow{g}+%
\overrightarrow{q}\right) \widetilde{U}(\overrightarrow{q}) 
\]
\[
\times e^{-in\theta _1(\overrightarrow{q})+i(N+n)\theta (\overrightarrow{p}+%
\overrightarrow{q})}\left[ \omega \left\{ \alpha \left( \frac{%
\overrightarrow{p}}{k\cdot p}\right) D_{1,n}^{\dagger }(\theta _1(%
\overrightarrow{q})-\theta (\overrightarrow{p}))-Z(1-\zeta
^2)D_{2,n}^{\dagger }\right\} -\Pi _0D_n^{\dagger }\right] 
\]
\[
\times \left. \frac{J_{(N+n)}\left( \alpha \left( \frac{\overrightarrow{p}+%
\overrightarrow{q}}{k\cdot p-\overrightarrow{k}\cdot \overrightarrow{q}}%
\right) ,-\frac{Z_1}2(1-\zeta ^2),\theta (\overrightarrow{p}+\overrightarrow{%
q})\right) }{\overrightarrow{q}^2+2\overrightarrow{\Pi }{\bf \cdot }%
\overrightarrow{q}-2n(k\cdot p-\overrightarrow{k}{\bf \cdot }\overrightarrow{%
q})+i0}\right\} 
\]
\begin{equation}
\times \delta \left( \Pi _0-\varepsilon _0-\omega N\right) .  \label{34}
\end{equation}
For hydrogen-like atoms with the charge number $Z_a$ the condition of the
Born approximation for the photoelectron scattering (in Coulomb field)

\begin{equation}
\frac{Z_ae^2}{\hbar v}\ll 1  \label{35}
\end{equation}
requires electron velocities $v\gg Z_a\alpha $, where $\alpha =e^2/\hbar
c=1/137$ is the fine-structure constant (it is assumed that $Z_a<<137,$ $%
\hbar $ and $c$ are restored for clarity). The photoelectron acquires such
velocities in the EM wave field at the intensities

\begin{equation}  \label{35a}
\xi \gg \frac{Z_a}{137}.
\end{equation}
As will be shown bellow the Eq. (\ref{35a}) is the condition of the Born
approximation in ATI process of hydrogen-like atoms taking into account the
photoelectron rescattering.

The initial bound state enters into Eq. (\ref{29}) through its momentum
space wave function $\widetilde{\Phi }\left( \overrightarrow{b}\right) $.
For hydrogen-like atom the bound state wave function is 
\begin{equation}
\Phi (\overrightarrow{\eta }{\bf )=}\frac{\exp (-\eta /a)}{\sqrt{\pi a^3}},
\label{36}
\end{equation}
where $a=a_0/Z_a$ ($a_0=1/me^2$ is the Bohr radius) (\ref{33}) and the
corresponding momentum space wave function has the following form: 
\begin{equation}
\widetilde{\Phi }\left( \overrightarrow{b}\right) =\frac{2^3(\pi a^3)^{1/2}}{%
\overrightarrow{b}^4a^4}.  \label{37}
\end{equation}
Note, that in Eq. (\ref{37}) it has been taken into account that $\left| 
\overrightarrow{b}\right| a\gg 1$ in accordance with the Born approximation.
Then the function $\widetilde{\Phi }\left( \overrightarrow{g}+%
\overrightarrow{q}\right) $ in the second term in figured brackets can be
replaced by the quantity $\delta (\overrightarrow{g}+\overrightarrow{q})/%
\sqrt{\pi a^3}$ because of the small contributions of the other terms in
expansion of $T_{i\rightarrow f}$ over parameter $\overrightarrow{g}^2a^2$
[see, e.g., \cite{48}] which will be shown bellow. Such a delta function can
be used to accomplish the integration over $\overrightarrow{q}$ in the
second term of the sum in the figured brackets of Eq. (\ref{34}).

For the scattering of a charged particle in the Coulomb field for which the
Fourier transform is 
\begin{equation}  \label{38}
\widetilde{U}(\overrightarrow{g}{\bf )=}\frac{4\pi }{am\overrightarrow{g}^2}%
{\bf ,}
\end{equation}
we have the following expression for the transition amplitude in the field
of arbitrary polarization of EM wave 
\[
T_{i\rightarrow f}=\frac{i2^4(\pi a)^{3/2}}{\sqrt{m\Pi _0}}\frac{(k\cdot p)}{%
\overrightarrow{g}^4a^4}\sum_{N=-\infty }^\infty \left\{ \left( N-Z(1+\zeta
^2)\right) e^{iN\theta (\overrightarrow{p})}J_N\left( \alpha \left( \frac{%
\overrightarrow{p}}{k\cdot p}\right) ,-\frac Z2(1-\zeta ^2),\theta (%
\overrightarrow{p})\right) \right. 
\]
\[
-\text{ }\frac{\omega \varepsilon _0\overrightarrow{g}^2}{m(k\cdot p)}%
\sum_{n=-\infty }^\infty \left( 2n-\alpha ^{\prime }(1+\zeta ^2)\right)
e^{-i(2n-N)\theta (\overrightarrow{p})} 
\]
\[
\times \left. \frac{\left\{ (\omega (2n-N)+\Pi _0)C_{N-2n}^{\dagger }+\omega
\alpha ^{\prime }(1-\zeta ^2)C_{2,N-2n}^{\dagger }\right\} J_n\left( \frac{%
-\alpha ^{\prime }(1-\zeta ^2)}2\right) }{m_{*}^2+\varepsilon
_0^2-2\varepsilon _0(\Pi _0+\omega (2n-N))}\right\} 
\]
\begin{equation}  \label{39}
\times \delta \left( \Pi _{0-}\varepsilon _0-\omega N\right) ,
\end{equation}
where $\alpha ^{\prime }$ is defined by Eq. (\ref{28b}) at $\overrightarrow{q%
}=-\overrightarrow{g}$and $\alpha ^{\prime }=e^2\overline{A}_0^2/4\omega
\varepsilon _0$ , then $J_n\left( \frac{-\alpha ^{\prime }(1-\zeta ^2)}2%
\right) $ is the ordinary Bessel function [ $J_{2n}\left( 0,x,0\right)
=J_n(x)$ (\ref{A6})], $C_s$ and $C_{2,s}$ are defined by the expressions 
\begin{equation}  \label{40}
C_s=J_s(\alpha (\overrightarrow{p}/kp)),(Z-\alpha ^{\prime })(1-\zeta
^2)/2,\theta (\overrightarrow{p})),
\end{equation}
and 
\[
C_{2,s}=\frac 12\left[ J_{s-2}(\alpha (\overrightarrow{p}/kp)),(Z-\alpha
^{\prime })(1-\zeta ^2)/2,\theta (\overrightarrow{p})e^{-i2\theta (%
\overrightarrow{p})}\right. 
\]
\begin{equation}  \label{41}
\left. +J_{s+2}(\alpha (\overrightarrow{p}/kp)),(Z-\alpha ^{\prime
})(1-\zeta ^2)/2,\theta (\overrightarrow{p})e^{i2\theta (\overrightarrow{p}%
)}\right] .
\end{equation}

To determine the differential probability of ATI it should be integrate the
expression (\ref{39}) over $\Pi _0$ according to (\ref{30}). In result we
have 
\[
\frac{dW_{i\rightarrow f}}{d\Omega }=\frac{2^4}{\pi ma^5}\sum_{N=N_0}^\infty 
\frac{\left( N-Z(1+\zeta ^2)\right) ^2(k\cdot \Pi )^2\left| \overrightarrow{%
\Pi }\right| }{\overrightarrow{g}^8} 
\]
\[
\times \left| \left\{ e^{iN\theta (\overrightarrow{\Pi })}J_N\left( \alpha
\left( \frac{\overrightarrow{\Pi }}{k\cdot \Pi }\right) ,-\frac Z2(1-\zeta
^2),\theta (\overrightarrow{\Pi })\right) \right. \right. 
\]
\[
+\frac{\overrightarrow{g}^2}{2m\left( N-Z(1+\zeta ^2)\right) (k\cdot \Pi )}%
\sum_{n=-\infty }^\infty e^{-i(2n-N)\theta (\overrightarrow{\Pi })}J_n\left( 
\frac{-\alpha ^{\prime }(1-\zeta ^2)}2\right) 
\]
\begin{equation}
\left. \times \left. \left( (\varepsilon _0+2n\omega )C_{N-2n}^{\dagger
}+\omega \alpha ^{\prime }(1-\zeta ^2)C_{2,N-2n}^{\dagger }\right) \right\}
\right| ^2,  \label{42}
\end{equation}
where $\overrightarrow{g}$ is the three-vector 
\begin{equation}
\overrightarrow{g}=\overrightarrow{\Pi }-N\overrightarrow{k},\ \left| 
\overrightarrow{\Pi }\right| =\sqrt{(\varepsilon _0+\omega N)^2-m_{*}^2}.
\label{43}
\end{equation}
The number $N_0$ over which is carried out summation in Eq. (\ref{42}) is
defined from the energy conservation law of ATI process: $N_0=\left\langle
(m_{*}-\varepsilon _0)/\omega \right\rangle $ .

The first term in the figured brackets of Eq. (\ref{42}) corresponds to
result of the KFR approximation. And the second term shows the dependance of
ATI probability on the ejected photoelectron stimulated bremsstrahlung (SB)
probability, i.e., it takes into account the rescattering process.

\section{Probability of ATI process for the circular and linear polarization
of EM wave}

The state of photoelectron in the field of a strong EM wave and consequently
the ionization probability essentially depends on the polarization of the
wave [nonlinear effect of intensity conditioned by the impact of strong
magnetic field]. Thus, for the circular polarization the relativistic
parameter of the wave intensity $\xi ^2=const=\xi _0^2$ and the longitudinal
velocity of the electron in the wave $v_{II}=const$ (eliminating this
inertial motion - in the framework connecting with the electron - we have
the uniform rotation in the polarization plane with the wave frequency $%
\omega $), meanwhile for the linear one $\xi ^2=\xi _0^2\cos ^2\varphi $ and 
$v_{II}$ oscillates with the frequencies of all wave harmonics $n\omega $
corresponding to strongly unharmonic oscillatory motion of photoelectron.
The later leads to principally different behavior of ionization process and
corresponding formulas depending on the polarization of strong wave.
Therefore, we shall consider the cases of circular and linear polarizations
of EM wave field separately.

From the Eq. (\ref{42}) for the circularly polarized wave ($\zeta =1$) in
the first Born approximation by ionized atom potential we obtain the next
formula for differential probability of ATI process 
\[
\frac{dW_{i\rightarrow f}}{d\Omega }=\frac{2^4}{\pi ma^5}\sum_{N=N_0}^\infty 
\frac{\left( N-2Z\right) ^2(k\cdot \Pi )^2\left| \overrightarrow{\Pi }%
\right| }{\overrightarrow{g}^8}J_N^2\left( \alpha \left( \frac{%
\overrightarrow{\Pi }}{k\cdot \Pi }\right) \right) 
\]
\begin{equation}  \label{44}
\times \left\{ 1+\frac{\overrightarrow{g}^2}{2\left( N-2Z\right) (k\cdot \Pi
)}\right\} ^2.
\end{equation}
As is seen from this formula in contrast to the case of another
polarizations the differential probability of ATI process is defined by
ordinary Bessel function instead of the function $J_n(u,v,\triangle )$ and
the sum over $n$ vanishes. The latter corresponds the above mentioned fact
that for the circular polarization the parameter of intensity of the wave $%
\xi ^2=const$ and effect of intensity of strong wave is appeared in the form
of constant renormalization of the characteristic parameters of the
interacting system.

Let us estimate the contribution of photoelectron rescattering in the
probability of ATI process that is the second term in the figured brackets
in Eq. (\ref{44}). The latter is 
\begin{equation}
\frac{\overrightarrow{g}^2}{2\left( N-2Z\right) (k\cdot \Pi )}\simeq 1
\label{45}
\end{equation}
for the most probable number of absorbed photons at which the Bessel
function has the maximum value. So, the rescattering effect has the same
order as the probability of the direct transition in SFA. It is followed to
note that the derivations rely upon the SFA (e.g., \cite{38}) are expected
to become more accurate at high intensity EM field. However, the prediction
of SFA regarding to rescattering effect in high intensity EM field, i.e.,
for relativistic photoelectron- according to which the rescattering will be
negligible small in relativistic domain with the increasing the wave
intensity \cite{38}- is not true, specially for Coulomb field, as we have
obtained significant contribution even in the Born approximation when the
impact of scattering potential is the smallest. Indeed, beyond the scope of
the Born approximation the contribution of rescattering effect in ATI
process will be more considerable (for instant, in the above considered GEA
for scattering potential).

In the scope of current approximation $\xi \gg Z_a/137$ the explicit
analytic formulas for total ionization rate can be obtained utilizing
properties of Bessel function. At the condition (\ref{35a}) the argument of
Bessel function $X(N)>>1$ and always $X<N$. Therefore the terms with $N>>1$
and $N\sim X$ give the main contribution in the sum (\ref{44}). Besides, in
this limit one can replace the summation over $N$ with integration and
approximate the Bessel function by Airy one

\begin{equation}  \label{46}
J_N(x)\simeq \left( \frac 2N\right) ^{1/3}Ai\left[ \left( \frac N2\right)
^{2/3}\left( 1-\frac{x^2}{N^2}\right) \right]
\end{equation}

Turning to spherical coordinates, we carry out the integration over the $%
\varphi $ since there is azimutal symmetry with respect to the direction $%
\overrightarrow{k}$ (the $OZ$ axis) and for ionization rate we have

\[
W_{i\rightarrow f}=\frac{2^5}{ma^5}\int_0^\pi \sin \theta d\theta
\int_{N=N_0}^\infty \left( \frac 2N\right) ^{2/3}\frac{(N-2Z)^2(k\cdot \Pi
)^2\left| \overrightarrow{\Pi }\right| }{\overrightarrow{g}^8}Ai^2\left[
y(N,\theta )\right] 
\]
\begin{equation}  \label{47}
\times \left\{ 1+\frac{\overrightarrow{g}^2}{2\left( N-2Z\right) (k\cdot \Pi
)}\right\} ^2.
\end{equation}

where 
\begin{equation}  \label{48}
y(N,\theta )=\left( \frac N2\right) ^{2/3}\left[ 1-\frac{\alpha ^2\left( 
\frac{\overrightarrow{\Pi }}{k\cdot \Pi }\right) }{N^2}\right]
\end{equation}

The $y(N,\theta )$ has a minimum as a function of $N$ and $\theta $ and
since Airy function exponentially decreases with increasing argument one can
use Laplace method ( method of the steepest descent) in order to carry out
the integration as over $N$ as well as over $\theta $. The extremum points
of the function $y(N,\theta )$, i.e., the most probable values of $N$ and $%
\theta $ are

\begin{equation}  \label{49}
N_m=\frac{m_{*}^2-\varepsilon _0^2}{\varepsilon _0\omega }\simeq \frac m%
\omega \xi ^2;\qquad \cos \theta _m=\frac{\left| \overrightarrow{\Pi }%
(N_m)\right| }{\Pi _0(N_m)}
\end{equation}

and 
\begin{equation}
y_m=y(N_m,\theta _m)=\frac{2^{1/3}E_B}{N_m^{1/3}\omega }=\left( \frac{F_{at}%
}{2F_0}\right) ^{2/3},  \label{50}
\end{equation}
where $F_0$ and $F_{at}=Z_a^3m^2e^5$ are wave and atomic electric field
strengths. At $N=N_m$ and $\theta =\theta _m$ we have a peak for angular and
energetic distribution. Let us note that the contribution of the
rescattering effect to the angular distribution of the photoelectrons is
nonessential.

For $y_m<<1$ when the wave electric field strength much exceeds the atomic
one: $F_0>>F_{at}$ the main contribution in the integral give the areas 
\begin{equation}
\delta \theta \simeq (N_m/2)^{-1/3}/\sqrt{1+\xi ^2}\text{and }\delta N\simeq
2(N_m/2)^{2/3}  \label{50a}
\end{equation}
(angular and energetic widths of the peak) and for ionization rate we have
an explicit formula which expresses directly the dependence upon the wave
intensity

\begin{equation}
W_{i\rightarrow f}=\frac{2^{7/3}}{3^{4/3}\Gamma ^2(2/3)}\pi \omega \left( 
\frac \omega {E_B}\right) ^3\left( \frac{F_{at}}{F_0}\right) ^{11/3}\cdot
\label{51}
\end{equation}

For $y_m>>1$ or $F_0<<F_{at}$ (so called tunneling regime of ionization) we
shall use the following asymptotic formula for Airy function

\begin{equation}
Ai(x)\simeq \frac 1{2\sqrt{\pi }}x^{-1/4}\exp \left( -\frac{2x^{3/2}}3\right)
\label{52}
\end{equation}
and applying Laplace method we have

\begin{equation}
W_{i\rightarrow f}=2\omega \left( \frac \omega {E_B}\right) ^3\left( \frac{%
F_{at}}{F_0}\right) ^3\exp \left\{ -\frac 23\frac{F_{at}}{F_0}\right\} \cdot
\label{53}
\end{equation}

Let's revert to the Born condition (\ref{35}) to substantiate the condition (%
\ref{35a}). As is shown above we have a peak for angular and energetic
distribution (\ref{44}) at $\theta _m$ and $N_m$ (\ref{49}), and the
electron mean velocity will be defined by these values 
\begin{equation}
v=\frac{\left| \overrightarrow{\Pi }(N_m)\right| }{\Pi _0(N_m)}\simeq \frac %
\xi {\sqrt{1+\xi ^2}}  \label{54}
\end{equation}
$.$Substituting (\ref{54}) into Eq.(\ref{35}) we have the condition of the
Born approximation in ATI process of hydrogen-like atoms (\ref{35a}).

Using the explicit analytic formulas for total ionization rate we can
conclude that at $N=N_m$ and $\theta =\theta _m$ we have a peak for angular
and energetic distribution which are given by Eq. (\ref{49}) with the
angular and energetic widths of the peak $\delta \theta $ and $\delta N$,
respectively (\ref{50a}).

In the case of linear polarization of the wave from Eq. (\ref{42}) we have 
\[
\frac{dW_{i\rightarrow f}}{d\Omega }=\frac{2^4}{\pi ma^5}\sum_{N=N_0}^\infty 
\frac{(N-Z)^2(k\cdot \Pi )^2\left| \overrightarrow{\Pi }\right| }{%
\overrightarrow{g}^8} 
\]
\[
\times \left\{ J_N\left( \alpha \left( \frac{\overrightarrow{\Pi }}{k\cdot
\Pi }\right) ,-\frac Z2\right) +\frac{\overrightarrow{g}^2}{2m(N-Z)(k\cdot
\Pi )}\sum_{n=-\infty }^\infty J_n\left( -\alpha ^{\prime }/2\right) \right. 
\]
\[
\times \left[ \left( \varepsilon _0+2n\omega \right) J_{N-2n}\left( \alpha
\left( \frac{\overrightarrow{\Pi }}{k\cdot \Pi }\right) ,\left( Z-\alpha
^{\prime }\right) /2\right) +\frac{\omega \alpha ^{\prime }}2\right. 
\]
\begin{equation}
\left. \left. \times \left( J_{N-2n-2}\left( \alpha \left( \frac{%
\overrightarrow{\Pi }}{k\cdot \Pi }\right) ,\left( Z-\alpha ^{\prime
}\right) /2\right) +J_{N-2n+2}\left( \alpha \left( \frac{\overrightarrow{\Pi 
}}{k\cdot \Pi }\right) ,\left( Z-\alpha ^{\prime }\right) /2\right) \right)
\right] \right\} ^2,  \label{55}
\end{equation}
where $J_n(u,v)$ is the real {\it generalized} Bessel function (e.g., see 
\cite{18}). As is seen from the formula (\ref{55}), in this case the total
probability of ATI process includes all intermediate transitions of
photoelectron trough the virtual vacuum states as well, corresponding the
emission and absorption of wave photons of number $-\infty <n<\infty $ (the
sum over $n$) in accordance with the above mentioned behavior of wave
intensity effect at linear polarization (strongly unharmonic oscillatory
motion of photoelectron).

Let us now consider the ATI process with the rescattering effect in
nonrelativistic limit since the theoretical treatments of this problem- the
main of those are Keldysh-Faisal-Reiss ansatz (\cite{16}-\cite{18}) - in
general have been carried out for nonrelativistic photoelectron when the
rescattering effect is neglected. In the pioneer result of Keldysh (\cite{16}%
) the rescattering of photoelectron on the potential of atomic remainder has
been approximately estimated and putted in the form of coefficient in the
ultimate formula for the ionization probability (for the wave fields much
smaller than atomic ones). Further the same approach has been made in \cite
{49} for relatively large wave fields up to the atomic ones. Besides, in the
existing nonrelativistic theory of ATI the gauge problem for description of
interaction with the wave field and different views concerning the role of
wave intensity in the dipole approximation have arose. For discussion of
these problems the special paper has been devoted (\cite{e}). Moreover, in
the scope of the same Keldysh-Faisal-Reiss ansatz the existence of
stabilization effect depends on the gauge of the wave field (\cite{9}). So
that, we shall consider the results of the present paper in the
nonrelativistic limit taking into account the photoelectron rescattering.

From the formula (\ref{44}) for the differential probability of ATI
transition rate in the case of circular polarization of EM wave in the
nonrelativistic limit we have 
\[
\frac{dW_{i\rightarrow f}^{nrel}}{d\Omega }=\frac{8\omega }\pi \left( \frac{%
E_B}\omega \right) ^{5/2}\sum_{N=N_0}^\infty \frac{(N-2z-E_B/\omega )^{1/2}}{%
(N-2z)^2}J_N^2\left( \vartheta \right) 
\]
\begin{equation}
\times \left[ 1+\frac{N-2z-E_B/\omega }{N-2z}\right] ^2,  \label{56}
\end{equation}
where 
\begin{equation}
\vartheta =\frac{e\overline{A}_0}{\omega m}\sqrt{\left( \overrightarrow{p}%
\cdot \widehat{\overrightarrow{e}}_1\right) ^2+\left( \overrightarrow{p}%
\cdot \widehat{\overrightarrow{e}}_2\right) ^2},  \label{57}
\end{equation}

$z=Z=Z_1=e^2\overline{A}_0^2/4m\omega ,$ and $N_0=\left\langle \left( 
\overrightarrow{p}^2/2m-E_B\right) /\omega +z\right\rangle $.

The corresponding condition of the Born approximation (Eq. (\ref{35a})) in
nonrelativistic limit is 
\begin{equation}
1\gg \xi \gg \frac{Z_a}{137}.  \label{58}
\end{equation}
The first term in the quadratic brackets of Eq. (\ref{56}) coincides with
the above-threshold ionization differential probability obtained in the SFA
for the nonrelativistic photoelectron \cite{18} without the rescattering
effect. According to the \cite{38} SFA is expected to become valid when the
ponderomotive potential $U_p=e^2\overline{A}_0^2/2m$ due to EM radiation
field larger than the ionization potential of the atom: $U_p\gg E_B$ and
consequently$\overrightarrow{p}^2/2m\gg E_B$ which is the condition of the
Born approximation. Then taking into account the scattering potential by
perturbation theory we obtain [(\ref{56})] that the contribution of the
photoelectron rescattering in the ATI probability (in the first order of the
Born approximation over the Coulomb potential) is of the order of the main
results of KFR ansatz. So, the neglecting of SB process for the
photoelectron in the Coulomb field of an atomic remainder is incorrect.

In the case of linear polarized EM wave from the formula (\ref{55}) we have
the differential probability of the ATI process in the nonrelativistic
domain 
\[
\frac{dW_{i\rightarrow f}^{nrel}}{d\Omega }=\frac{8\omega }\pi \left( \frac{%
E_B}\omega \right) ^{5/2}\sum_{N=N_0}^\infty \frac{(N-z-E_B/\omega )^{1/2}}{%
(N-z)^2}J_N^2\left( u,-\frac z2\right) 
\]
\begin{equation}
\times \left\{ 1+\frac{(N-z-E_B/\omega )}{(N-z)}\right\}   \label{59}
\end{equation}
where 
\begin{equation}
u=z^{1/2}\chi ,\ \qquad \chi =8^{1/2}\left( N-z-\frac{E_B}\omega \right)
^{1/2}\cos \theta   \label{60}
\end{equation}
in the polar coordinate system has been chosen for this case, $\theta $ is
the angle between the velocity vector of the emitted photoelectron and the
wave polarization vector.

\appendix 

\section{Definition of the Function $J_n(u,v,\triangle )$}

A function $J_n(u,v,\triangle )$ may be defined by 
\begin{equation}
J_n(u,v,\triangle )=(2\pi )^{-1}\int_{-\pi }^\pi d\theta \exp \left[ i\left(
u\sin (\theta +\triangle )+v\sin 2\theta -n(\theta +\triangle )\right)
\right]  \label{A1}
\end{equation}
or by an infinite series representation 
\begin{equation}
J_n(u,v,\triangle )=\sum_{k=-\infty }^\infty e^{-i2k\triangle
}J_{n-2k}(u)J_k(v).  \label{A2}
\end{equation}
We are perform two important theorems, which can be proved from Eq. (\ref{A1}%
): 
\begin{equation}
\sum_{n=-\infty }^\infty e^{in(\varphi +\triangle )}J_n(u,v,\triangle )=\exp
\left\{ i\left[ u\sin (\varphi +\triangle )+v\sin 2\varphi \right] \right\}
\label{A3}
\end{equation}
and 
\begin{equation}
\sum_{k=-\infty }^\infty J_{n\mp k}(u,v,\triangle )J_k(u^{\prime },v^{\prime
},\pm \triangle )=J_n(u\pm u^{\prime },v\pm v^{\prime },\triangle ).
\label{A10}
\end{equation}
An integration by parts in Eq. (\ref{A1}) yields to the relation 
\[
2nJ_n(u,v,\triangle )=u\left[ J_{n-1}(u,v,\triangle )+J_{n+1}(u,v,\triangle
)\right] 
\]
\begin{equation}
+2v\left[ e^{-i2\triangle }J_{n-2}(u,v,\triangle )+e^{i2\triangle
}J_{n+2}(u,v,\triangle )\right] .  \label{A4}
\end{equation}
From either Eq. (\ref{A1}) or (\ref{A2}) follows that 
\begin{equation}
J_n(u,0,\triangle )=J_n(u),  \label{A5}
\end{equation}
and 
\begin{equation}
J_n(0,v,\triangle )=\left\{ 
\begin{array}{c}
e^{-i\triangle n}J_{\frac n2}(v),\ n\text{ even} \\ 
0,\ n\text{ odd}
\end{array}
\right. \ .  \label{A6}
\end{equation}

\end{document}